\documentclass[twocolumn,showpacs,amsmath,amssymb,pra,superscriptaddress]{revtex4-1}

\usepackage{epsfig}
\usepackage{graphicx}
\usepackage{dcolumn}
\usepackage{color}
\usepackage{bm}
\usepackage{mathrsfs}
\usepackage{amssymb}

\begin{document}

\title {Decoherence-free subspace and disentanglement dynamics for two qubits in a common non-Markovian squeezed reservoir}

\author{Md. Manirul Ali}
\affiliation{Research Center for Applied Sciences, Academia Sinica,
Taipei 11529, Taiwan}
\author{Po-Wen Chen}
\affiliation{Department of Physics and Center for Theoretical Sciences,
National Taiwan University, Taipei 10617, Taiwan}
\author{Hsi-Sheng Goan}
\email{goan@phys.ntu.edu.tw}
\affiliation{Department of Physics and Center for Theoretical Sciences,
National Taiwan University, Taipei 10617, Taiwan}
\affiliation{Center for Quantum Science and Engineering,
National Taiwan University, Taipei 10617, Taiwan}

\date{\today}

\begin{abstract}
We study the non-Markovian entanglement dynamics of two qubits in a common 
squeezed bath. We see remarkable difference between the non-Markovian 
entanglement dynamics with its Markovian counterpart. 
We show that a non-Markovian decoherence free state is also decoherence free 
in the Markovian regime, but all the Markovian decoherence free states
are not necessarily decoherence free in the non-Markovian domain. 
We extend our calculation from squeezed vacuum bath to squeezed thermal bath, 
where we see the effect of finite bath temperatures on the entanglement 
dynamics.
\end{abstract}

\maketitle

\section{Introduction}

Entanglement is a remarkable feature of quantum mechanics, and its investigation is both of
practical and theoretical significance. 
It is also viewed 
as a basic resource for quantum information processing \cite{book}, like 
entanglement-based quantum cryptography \cite{secure}, quantum teleportation 
\cite{tele}, dense coding \cite{dense} and 
cluster-state quantum computation \cite{cluster_state}. 
Thus the real world success of these quantum information processing schemes
relies on the longevity of entanglement 
in multiparticle quantum states. 
Entanglement is also related to the basic issue of 
understanding the nature of nonlocality in quantum mechanics \cite{epr,bell}. However, a quantum system 
used in quantum information processing inevitably interacts with the surrounding environment, which induces 
the quantum world into the classical world \cite{deco1,deco2}. The presence of decoherence in communication 
channels and computing devices degrades the entanglement when the particles propagate or computation evolves. 
The coupling of the quantum system with its surroundings and the consequent decay of entanglement motivate 
important questions such as to understand its sources, and possibly to find ways to circumvent it through 
different types of controlled environments.

The dynamics of open quantum systems, however, may be rather involved, mostly due to the complex structure 
of the environment interacting with the quantum system. Generally, the nonunitary evolution of the reduced 
density matrix of the system is obtained after taking partial trace over bath variables. In this process, 
some approximations are often made 
in the derivation of a master equation for the systems reduced 
density matrix. The most important approximations \cite{deco1,weiss} are the weak coupling or {\it Born approximation} 
assuming that the coupling between the system and the reservoir is small enough to justify a perturbative approach, 
and the {\it Markov approximation} assuming that 
the correlation time of the reservoir is very short compared to the 
typical system response time
so that the reservoir 
correlation function is assumed to be $\delta$-correlated in time. 
Although, the use of Markovian approximation 
is justified in a large variety of quantum optical experiments where entanglement has been produced, one 
should notice that non-Markovian effects are crucial, e.g., for high-speed quantum communication where 
the characteristic time of the relevant system become comparable with the reservoir correlation time, 
or if the environment is structured with a particular spectral density, e.g. for quantum systems (channels) embedded 
in solid-state devices, where memory effects are typically non-negligible. In these cases, the dynamics can be
substantially different from the Markovian one. Due to their fundamental importance in quantum
information processing and quantum computation, non-Markovian quantum dissipative systems have
attracted much attention in recent years \cite{deco1,hu,liu,lorenz,prager,sabrina,Yuan,cao,bellomo,feng,cui,mazzola,Sinayskiy}, one of 
the main purposes of which in the long run is to engineer different types of 
(artificial) reservoirs, and couple
them to the system in a controlled way. 
The non-Markovian features of system-reservoir interactions 
have made great progress, 
but the theory is far from completion, in particular 
how different kinds of non-Markovian environments 
influence the systems, and the difference between Markovian and non-Markovian 
system evolutions are still a subject that demands further investigations.

If the environment would act on the various parties the same way it acts on single systems, one would expect
that a measure of entanglement, say the concurrence, would also decay exponentially in time. However, this is
not always the case. Recently Yu and Eberly \cite{eberly} showed that under certain conditions, the dynamics
could be completely different and the quantum entanglement of a bipartite qubit system may vanish in a finite
time. They called this effect ``entanglement sudden death (ESD)''. This phenomenon of entanglement sudden death 
has been extensively studied \cite{sudden} in the context of Markovian master equation.  
Entaglement dynamics for system of two qubits interacting with their local independent reservoir
is completely different from that when they interact with the same common reservoir. We see that
entanglement may not be destroyed by the interaction with the environment, 
and sometimes it persists at
whatever be the temperature of the bath. A common reservoir indirectly couples the qubits, and there have been 
some suggestions \cite{sugg} for creating as well as enhancing \cite{ali} entanglement between two or more 
parties by their collective interactions with a common environment. In the same vein, non-Markovian entanglement 
dynamics is fundamentally different from the Markovian one. With respect to the study of non-Markovian dynamics 
of two independent qubits, each locally interacting with 
its own reservoir, it was shown \cite{bellomo,feng} that although 
no interaction is present or mediated between the qubits, there is a revival of their entanglement following the ESD.
The backaction of the non-Markovian reservoir \cite{bellomo,feng} 
is responsible for revivals of entanglement 
after sudden death. The dynamics of entanglement in two independent 
non-Markovian channels is shown \cite{cui} to be 
oscillating at high temperatures, whereas in the Markovian channel, entanglement was shown to decay exponentially. 
Appearance of sudden death and sudden birth of entanglement was also discussed \cite{mazzola} in common structured 
reservoirs.

Different schemes have been derived to remove the effects produced by the environment, for example, quantum error 
correction, decoherence-free subspace (DFS), dynamical decoupling and quantum Zeno effect. In the presence of the
environment, the DFS is a set of all states which is not affected at all by the interaction with the bath. There were 
some proposals related to the use of DFS as the memory space for storing the quantum information \cite{dfs}. Recently, 
decoherence free entanglement was studied for two two-level systems interacting with a common squeezed vacuum bath 
\cite{orszag1}. The Markovian entanglement dynamics of two two-level atoms that interact with a common squeezed 
vacuum reservoir was extensively studied \cite{orszag2}. The phenomenon of sudden death and revival of entanglement 
was investigated for the initial states that are very close to (as well as far from) the Markovian DFS. It was 
claimed that for states belonging initially to the DFS plane, the phenomenon of entanglement sudden death never
occurs. However, if the initial state is away from the DFS plane, the sudden death shows up, followed by sudden 
revival of entanglement. 

Several proposals to physically realize the squeezed reservoir were
put forward in the literature, the simplest of which consists in considering
a two-level atom immersed in a squeezed multimode radiation field
\cite{gardiner1,scully1,expt1}. Parkins {\it et al.} \cite{parkins1993} showed
how a two-level system can be coupled to an almost ideal squeezed vacuum by assuming
an atom strongly interacting with a cavity field which is illuminated by finite-bandwidth
squeezed light. It was also shown \cite{vitali1} how a squeezed environment can be 
obtained by means of a suitable feedback of the output signal corresponding to a
quantum-nondemolition (QND) measurement of suitable quadrature
operators.
In Ref.~\cite{lutken}, the authors showed how to mimic the interaction of a
two-level system with a squeezed reservoir by using a four-level atom interacting with
circularly polarized laser fields. Assuming a strong decay of the two most excited levels,
it was shown that the dynamics of the two ground atomic states is effectively similar to
that of a two-level system interacting with a squeezed reservoir. A recent theoretical
study \cite{werlang} was made to generate squeezed reservoir for a two-level system by
engineering the Hamiltonian of a ${\Lambda}$-type three-level atom interacting with a single cavity mode and
laser fields with suitable intensity and detuning. Another experimental proposal
\cite{parkins2003} for two two-state atoms in a common
squeezed reservoir was made employing quantum-reservoir engineering to controllably
entangle the internal states of two atoms trapped in a high-finesse optical cavity.
Using laser and cavity fields to drive two separate Raman transitions between stable
atomic ground states, a system, corresponding to a pair of two-state
atoms coupled collectively to a squeezed reservoir, could be realized.

In this paper, we will analyze the entanglement dynamics for two qubits interacting with 
a common squeezed reservoir in the non-Markovian regime. We see multiple cycles of entanglement sudden death and 
revival in the non-Markovian case, showing striking difference between the Markovian and non-Markovian entanglement 
dynamics. We extend our result for the finite temperature case where we see the non-Markovian entanglement oscillations 
gradually decreases as one increases the temperature. Finite temperature of the bath accelerates the phenomenon of 
ESD in general. We show that the Markovian decoherence free states remains invariant under finite temperature of the 
bath, and all states in the Markovian DFS plane is not necessarily decoherence free in the non-Markovian regime. Interestingly, the singlet state (which satisfy a more general DFS condition) is found to be decoherence free 
both in the Markov and non-Markov regime, and is also found to be robust against finite bath temperature. In Sec.~\ref{sec:model}, we describe our model and present the non-Markovian master equation. In Sec.~\ref{sec:difference}
we discuss the difference of DFS and entanglement dynamics between 
the Markovian and non-Markovian cases. 
In Sec.~\ref{sec:numerical} we present the numerical results and discuss 
our main observations, ending it with some concluding remarks 
in Sec.~\ref{sec:conclusion}.

\section{Model and quantum master equations}\label{sec:model}

We consider a pair of two-level atoms (two qubits) coupled to a common 
non-Markovian thermal squeezed reservoir. The 
microscopic Hamiltonian of the system plus reservoir is given by
\begin{eqnarray}
H= \omega_0 \left( \sigma^1_{+} \sigma^1_{-} + \sigma^2_{+} \sigma^2_{-} \right) 
+ \sum_k \omega_k {b_k}^{\dagger} {b_k} + H_{I},
\label{hamilton}
\end{eqnarray}
where the interaction Hamiltonian $H_{I}$ has the form
\begin{eqnarray}
H_{I}=\sum_k  g_k ~S_{+} b_k + g^{\ast}_{k}~ S_{-} {b_k}^{\dagger}.
\label{interH}
\end{eqnarray}
Here $S_{+}=\sigma_{+}^1+\sigma_{+}^2$ and
$S_{-}=\sigma_{-}^1 +\sigma_{-}^2$ are collective
raising and lowering operators for the two-qubit system with $\sigma_{+}^{i}=|1_i\rangle\langle0_i|$,~~
$\sigma_{-}^{i}=|0_i\rangle\langle1_i|$, where $|1_i\rangle$ and $|0_i\rangle$ are up and down states 
of the {\it i}th qubit, respectively. Let us now proceed for the master equation for this two-qubit 
system interacting with a common squeezed thermal bath according to the Hamiltonian (\ref{hamilton}). 
We assume the factorized initial system-reservoir state with the initial state of the reservoir as a 
squeezed thermal equilibrium state given by 
\begin{eqnarray}
\rho_R(0) = \prod_k U(r_k,\theta_k)~ \rho_{th} ~U^{\dagger}(r_k,\theta_k),
\end{eqnarray}
\begin{eqnarray}
U(r_k,\theta_k) = \exp \left( \frac{1}{2} \xi_k^{\ast} b_k^2 - \frac{1}{2} \xi_k {b_k^{\dagger}}^2 \right),
\end{eqnarray}
\begin{eqnarray}
\rho_{th} =  \frac{\exp \left(-\beta \omega_k {b_k}^{\dagger} {b_k} \right)}{Tr~ \exp \left(-\beta \omega_k {b_k}^{\dagger} {b_k} \right)},
\end{eqnarray}
where $\beta=1/{KT}$ with $K$ being the Boltzman constant and $T$ being the temperature, and we have introduced the unitary squeeze \cite{deco1} operator $U(r_k,\theta_k)$ with 
$\xi_k = r_k e^{i \theta_k}$.

The non-Markovian 
master equation in the interaction picture 
for the two-qubit reduced 
density matrix $\rho(t)$ in the Born approximation 
can be calculated and written as:
\begin{eqnarray}
\nonumber
\frac{\partial \rho}{\partial t} &=&
\Delta(t)~ \{ S_{+} \rho S_{-} - \rho S_{-} S_{+} \}
+ \Delta^{\ast}(t) \{S_{+} \rho S_{-} - S_{-} S_{+} \rho \}\\
\nonumber
&+& \mu(t)~ \{S_{-} \rho S_{+} - S_{+} S_{-} \rho \} 
+ \mu^{\ast}(t)~ \{S_{-} \rho S_{+} - \rho S_{+} S_{-} \}\\
\nonumber
&+& \alpha(t) ~ \{ 2 S_{+} \rho S_{+} - S_{+} S_{+} \rho
- \rho S_{+} S_{+} \} \\
&+& \alpha^{\ast}(t)~ \{ 2 S_{-} \rho S_{-} - S_{-} S_{-} \rho
- \rho S_{-} S_{-} \}.
\label{nonmarkov}
\end{eqnarray}
The time dependent coefficients appearing in the master equation are respectively given by
\begin{eqnarray}
\Delta(t)& =& \int_0^{t}dt_1 \int_0^{\infty} d\omega ~J(\omega) N(\omega) e^{i(\omega_0-\omega)(t-t_1)}, \\
\mu(t)& = &\int_0^{t}dt_1 \int_0^{\infty} d\omega ~J(\omega) \left[1+ N(\omega)\right] e^{i(\omega_0-\omega)(t-t_1)}, \\
\alpha(t)& = &\int_0^{t}dt_1 \int_0^{\infty} d\omega ~J(\omega) M(\omega) e^{i(\omega_0-\omega)(t+t_1)}, 
\end{eqnarray}
where
\begin{eqnarray}
N(\omega)&=&n(\omega) \left[ \cosh^2 r + \sinh^2 r \right] + \sinh^2 r, 
\label{Nw} \\
M(\omega)&=&-\cosh r ~\sinh r ~e^{i\theta} \left[2 n(\omega) +1 \right]
\label{Mw}
\end{eqnarray}
with the frequency independent resonant squeeze parameter $r$ and the resonant phase $\theta$,
and we have written $n(\omega)=1/\left[\exp(\beta \omega)-1 \right]$ for the Planck distribution.
The master equation (\ref{nonmarkov}) is valid for arbitrary temperature 
(provided that the Born approximation still holds) and the squeezed vacuum 
reservoir case is just the zero-temperature limit of it. 
In the zero-temperature limit, $n(\omega)=0$,
so, $N(\omega)=N=\sinh^2(r)$, and $M(\omega)=-\cosh(r) \sinh(r)~ e^{i \theta}$=
$-M~e^{i \theta}$, where $M=\sinh(r) \cosh(r)=\sqrt{N(N+1)}$. The non-Markovian character 
is contained in the time-dependent coefficients, which contain the information about 
the system-reservoir correlations. In the previous equations, 
$J(\omega)=\sum_{k} |g_k|^2 \delta(\omega-\omega_k)$ is the spectral 
density characterizing the bath, where 
the index $k$ labels the different field mode of the reservoir 
with frequency $\omega_k$. 
We may consider any form of the reservoir spectral density.
But for simplicity, here we consider an Ohmic squeezed bath \cite{hu,liu} with the spectral density given by
\begin{eqnarray}
J(\omega)= \Gamma \omega \exp \left(- \omega^2 / {\omega_c}^2 \right),
\end{eqnarray}
where $\omega$ is the frequency of the bath and $\omega_c$ is the high-frequency cutoff and
$\Gamma$ is a dimensionless constant characterizing the interaction strength to the environment.
For finite temperature Markovian case, $n(\omega)=n(\omega_0)$. 
Since we consider two 
qubits in a common bath, they will have to be quite
near so that the interatomic 	 
separation is much smaller than a typical wavelength of the bath,
i.e., the length scale of the resonant wavelength $\lambda_0=\hbar 
c/\omega_0$ in the model \cite{Ficek02}, 
where $c$ is the wave speed of the bath. 	 
Thus we do not consider here the effect of qubit size, also the spatial
dependence on  
qubit-environment coupling strength \cite{Ficek02,space}
($g_k$ is assumed to be position-independent). We also assume that there is no direct interaction between the 
qubits except the indirect coupling through the common environment.

\section{Difference between Markovian and non-Markovian dynamics}
\label{sec:difference}

Our next aim is to show the difference between the Markovian and non-Markovian entanglement dynamics 
for this system-reservoir model. The Markovian master equation, in the interaction picture, for two 
two-level systems interacting with a broadband squeezed vacuum bath (at
zero temperature) is well studied 
\cite{orszag1,orszag2,orszag3} and is given by 
\begin{eqnarray}
\nonumber
\frac{\partial \rho}{\partial t} &=& \frac{1}{2} \gamma (N+1) 
\left(2 S_{-} \rho S_{+} - S_{+} S_{-} \rho - \rho S_{+} S_{-} \right)\\
\nonumber
&+& \frac{1}{2} \gamma N \left(2 S_{+} \rho S_{-} - S_{-} S_{+} \rho -\rho S_{-} S_{+} \right)\\
\nonumber
&-& \frac{1}{2} \gamma M e^{i \theta}  \left(2 S_{+} \rho S_{+} - S_{+} S_{+} \rho - \rho S_{+} S_{+} \right)\\
&-& \frac{1}{2} \gamma M e^{-i \theta} \left(2 S_{-} \rho S_{-} - S_{-} S_{-} \rho - \rho S_{-} S_{-} \right)
\label{markov}
\end{eqnarray}
\noindent
where $\gamma=\pi \Gamma \omega_0$. Using this Markovian master equation (\ref{markov}), the entanglement dynamics 
(the phenomenon of sudden death and revival of entanglement) of a pair of two-level atoms has been extensively 
studied \cite{orszag2}. The DFS for this model (Markov approximation, broadband squeezed vacuum) is found in 
Ref.~\cite{orszag1}, and we call it Markovian DFS. 

The main result of the present paper will rotate around the discussion of 
DFS and entanglement dynamics according to the
Markovian master equation (\ref{markov}) and 
its non-Markovian counterpart (\ref{nonmarkov}), showing striking 
difference between them. 
This Markovian master equation (\ref{markov}) can be written in an explicit 
Lindblad form 
using only one Lindblad operator \cite{orszag1}:
\begin{eqnarray}
\frac{\partial \rho}{\partial t}= \frac{\gamma}{2} \left(2 L \rho L^{\dagger} -\rho L^{\dagger} L - L^{\dagger} L \rho \right),
\end{eqnarray}
where
\begin{eqnarray}
L=\sqrt{N+1}~ S_{-} - \sqrt{N} \exp\{i\theta\}~ S_{+}.
\end{eqnarray}
In this case, the DFS \cite{lider} is composed of all 
eigenstates of $L$ with zero eigenvalues. The two 
orthogonal vectors in the DFS plane for this Markovian evolution 
are \cite{orszag1}
\begin{eqnarray}
|\phi_1\rangle&=& \frac{1}{\sqrt{N^2 + M^2}} \left( N |11\rangle + M e^{-i\theta} |00\rangle \right),
\label{phi1}\\
|\phi_2\rangle&=&\frac{1}{\sqrt{2}} \left(|01\rangle - |10\rangle \right).
\label{phi2}
\end{eqnarray}
One can also define the states $|\phi_3\rangle$ and $|\phi_4\rangle$ orthogonal to $\{|\phi_1\rangle, |\phi_2\rangle \}$ plane:
\begin{eqnarray}
|\phi_3\rangle&=&\frac{1}{\sqrt{2}} \left(|01\rangle + |10\rangle \right),
\\
|\phi_4\rangle&=& \frac{1}{\sqrt{N^2 + M^2}} \left( M |11\rangle - N e^{-i\theta} |00\rangle \right).
\end{eqnarray}

It is important to mention here that the state $|\phi_1\rangle$, 
Eq.~(\ref{phi1}), is a decoherence free state only for the Makovian evolution
(\ref{markov}), but for a general non-Markovian dynamics $|\phi_1\rangle$ is not decoherence free. A more general 
discussions on DFS condition was discussed in Refs.\cite{lider,zanardi}. It is important to note that decoherence is the result
of the entanglement between system and bath caused by the 
interaction term $H_I$, Eq.~(\ref{interH}), 
of the Hamiltonian (\ref{hamilton}). 
In other words, if $H_I=0$ then system and bath are decoupled and evolve independently and unitarily under their
respective Hamiltonians $H_S$ and $H_B$. Clearly, then, a sufficient condition for decoherence free dynamics is 
that $H_I=0$. However, since one cannot simply switch off the system-bath interaction, in order to satisfy this 
condition, it is necessary to look for special subspace 
(say, ${\cal H}$) of the full system Hilbert space such that  
the system evolves in a completely unitary fashion on that 
subspace ${\cal H}$. As shown first by Zanardi and 
Rasetti \cite{zanardi}, such a subspace is found by assuming that there exists a set of degenerate eigenvectors 
of the system coupling operators $S_{\pm}$ in the 
system-reservoir interaction Hamiltonian.
In our case, focusing on the form of interaction Hamiltonian $H_I$ given by 
Eq.~(\ref{interH}), it is clear that the DFS is made up
of those states $|\psi\rangle$ satisfying \cite{lider} 
\begin{eqnarray}
S_{\pm}|\psi\rangle = 0.
\label{dfs}
\end{eqnarray}
The singlet state, Eq.~(\ref{phi2}), is one which 
satisfies (\ref{dfs}) with vanishing total angular momentum. 
Hence the singlet
state $|\psi\rangle=|\phi_2\rangle=(|01\rangle - |10\rangle)/{\sqrt 2}$~ is a decoherence free state for this type
of Hamiltonian. 
This is why the singlet state $|\phi_2\rangle$ is decoherence free both 
for Markovian (\ref{markov}) and non-Markovian (\ref{nonmarkov}) dynamics 
at any temperature. 
This result is also confirmed by the numerical calculations shown in 
Sec.~\ref{sec:numerical}. 

We note here that the qualitative characteristics of the 
non-Markovian entanglement dynamics with oscillations and 
sudden deaths and revivals (shown in Sec.~\ref{sec:numerical}) 
may not be specific to the squeezed reservoir, 
but the quantitative
feature of this dynamics will depend on various properties of the
reservoir, such as
spectral density, squeezing, and temperature. 
Nevertheless, it is important to mention here that
the Markovian master equation for two two-level system interacting with a common
({\it unsqueezed}) heat bath is
\begin{eqnarray}
\nonumber
\frac{\partial \rho}{\partial t} &=& \frac{1}{2} \gamma \left(N+1\right)
\left( 2 S_{-} \rho S_{+} - S_{+} S_{-} \rho - \rho S_{+} S_{-} \right)\\
&+& \frac{1}{2} \gamma N
\left( 2 S_{+} \rho S_{-} - S_{-} S_{+} \rho - \rho S_{-} S_{+} \right).
\end{eqnarray}
In this case, the DFS is composed of common eigenstates of the Lindblad operators
$L_{\pm}=S_{\pm}$ with zero eigenvalues. The singlet state
$|\phi_2\rangle=(|01\rangle - |10\rangle)/{\sqrt 2}$ satisfies this condition.
So, the singlet state $|\phi_2\rangle$ is the only state which is decoherence
free for an unsqueezed common reservoir. Interestingly, we see that for an unsqueezed
reservoir, DFS calculated from the Lindblad operators $L_{\pm}=S_{\pm}$ and that
obtained from the system-bath interaction (Zanardi and Rasetti criterion 
\cite{lider,zanardi}) Hamiltonian $H_I$, are the same. Hence there will be no difference 
between the Markovian and non-Markovian entanglement dynamics for this DFS state. On the
other hand, for the squeezed reservoir, one can have infinitely many Markovian DFS states
$|\phi_1\rangle$, Eq.~(\ref{phi1}), just by varying continuously the
squeeze parameters $\theta$
and $r$. In this case, squeezing plays an important role in showing difference between
the Markovian and non-Markovian entanglement dynamics for these Markovian DFS states. 
This is quantitatively shown in Fig.~\ref{prothom} in the next
section. Now, in order to study the  
sudden death and revival of entanglement of two qubits in this common 
squeezed bath (both in the Markovian 
and non-Markovian regimes), we consider as initial states of the 
form \cite{orszag1,orszag2}
\begin{eqnarray}
|\Psi_1 \rangle = \epsilon |\phi_1\rangle + \sqrt{1-\epsilon^2} |\phi_4\rangle,
\end{eqnarray} 
\begin{eqnarray}
|\Psi_2 \rangle = \epsilon |\phi_2\rangle + \sqrt{1-\epsilon^2} |\phi_3\rangle,
\end{eqnarray}
where $\epsilon$ is a variable amplitude of one of the states belonging to the Markovian DFS plane $\{|\phi_1\rangle, |\phi_2\rangle \}$. 
We would like to study the effect of varying $\epsilon$ ($0 \le \epsilon \le 1$) on the sudden death and revival of entanglement 
for these initial states. We calculate the time-evolved two-qubit density matrix for the initial states $|\Psi_1\rangle$ and $|\Psi_2\rangle$ using 
the Markovian (\ref{markov}) as well as the non-Markovian (\ref{nonmarkov}) master equations. The various components of the 
time-dependent density matrix depend on the initial states as well as on the squeezing parameters.

\section{Numerical results}\label{sec:numerical}

\begin{figure}
\includegraphics[width=\linewidth]{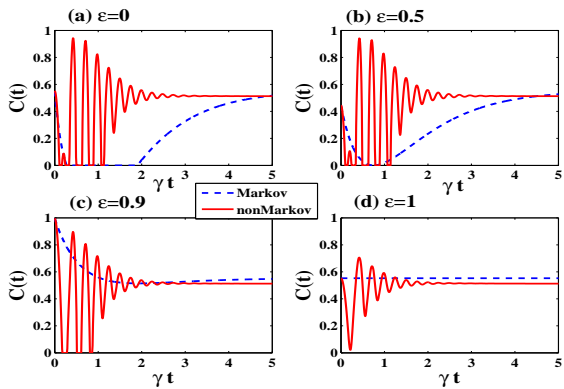}
\caption{(Color online) Non-Markovian (solid lines) and Markovian (dashed lines) time 
evolution of the concurrence for $|\Psi_1\rangle$ as an initial state with varying 
$\epsilon$: (a) $\epsilon=0$, (b) $\epsilon=0.5$, (c) $\epsilon=0.9$, and (d) $\epsilon=1$. 
The values of the other parameters are $r=0.31$, $\theta=0$, $\gamma=1$, and $\omega_c=
\omega_0=1$. The number of ESD and revivals in the non-Markovian case increases. The state 
$|\phi_1\rangle$ ($|\Psi_1\rangle$ with $\epsilon=1$) no longer remains decoherence free 
for the non-Markovian evolution although it is decoherence free in the Markovian case.}
\label{Mani1}
\end{figure}

We calculate numerically the time evolution of the density matrix according to Eqs.~(\ref{markov}) 
and (\ref{nonmarkov}) and we choose the Wootters entanglement measure \cite{wootters}, the 
concurrence $C(t)$, defined for the time-evolved two-qubit density matrix $\rho(t)$ as
\begin{eqnarray}
C(t)=\max(0, \sqrt{\lambda_1}-\sqrt{\lambda_2}-\sqrt{\lambda_3}-\sqrt{\lambda_4}),
\end{eqnarray}
where $\lambda_1$, $\lambda_2$, $\lambda_3$, $\lambda_4$ are the eigenvalues of the matrix
$\rho_c=\rho(t)(\sigma_y\otimes\sigma_y) {\rho}^{\ast}(t) (\sigma_y\otimes\sigma_y)$ in descending
order. 
The entanglement dynamics (in both the Markovian and 
non-Markovian regimes) is shown in  
Fig.~\ref{Mani1} for two qubits initially in 
the state $|\Psi_1\rangle$. The state $|\Psi_1\rangle$ is a 
superposition of two states $|\phi_1\rangle$ (belonging to the Markovian DFS) and its orthogonal 
$|\phi_4\rangle$. We vary $\epsilon$ between $0$ and $1$ for fixed values of the parameters $r=0.31$ and
$\theta=0$ as in Ref.\cite{orszag2}. 
Let us recapitulate the observation made in Ref.\cite{orszag2} about the Markovian dynamics of 
entanglement for the state $|\Psi_1\rangle$ in the parameter interval $0 \le \epsilon < 0.5$ [see also the curve in dashed line in Fig.~\ref{Mani1}(a)], where it was 
shown that the initial entanglement decays to zero in a finite time $t_d$. 
Then after a finite period of time 
during which the concurrence stays null, 
it revives at a later time $t_r$, then reaching asymptotically 
its steady-state value. For this Markovian case, 
it was also observed that this death and revival cycle happens 
only once for the initial state $|\Psi_1\rangle$. 
For $\epsilon=0.5$, the entanglement dies and revives 
simultaneously and eventually goes to its steady-state value 
[see also the curve in dashed line in Fig.~\ref{Mani1}(b)]. 
For $0.5 \le \epsilon < 1$, no entanglement 
sudden death was found \cite{orszag2} in the Markovian dynamics
[see also the curve in dashed line in Fig.~\ref{Mani1}(c)]. 
Finally, when $\epsilon =1$, $|\Psi_1\rangle=|\phi_1\rangle$
is a decoherence-free state so that the concurrence remains 
constant with time [see also the curve in dashed line in Fig.~\ref{Mani1}(d)]. 
This was the picture for the Markovian master 
equation (\ref{markov}). Now we compare this Markovian case with 
the non-Markovian entanglement dynamics according to 
Eq.(\ref{nonmarkov}) for the initial state $|\Psi_1\rangle$.

In the non-Markovian case for $\epsilon=0$, 
we see from the curve in solid line in Fig.\ref{Mani1}(a) 
that the initial entanglement decays to zero 
in a finite time showing suddent death of entanglement and then it revives again, following four successive death 
and revival, and finally the steady state value of the concurrence is reached at large times. Whereas in the case 
of Markovian dynamics, we see only one death and revival. It is also important to note that the entanglement sudden 
death occurs much faster in the non-Markovian case compared to the 
Markovian one. The time gap between adjacent death 
and revival is small (that is the rate at which the death and revival occurs are very fast) compared to its Markovian 
counterpart. The striking difference shown in Fig.\ref{Mani1}(a) between the Markov and non-Markov entanglement dynamics is 
that in the non-Markovian case, the entanglement is nonzero (showing three revival cycle) in a time window when the 
concurrence remains null in the case of Markovian dynamics. Similar multiple death and revival cycle is observed for 
the state with $\epsilon=0.5$ in the non-Markovian case showing a clear departure from the Markovian dynamics. When 
$0.5 \le \epsilon < 1$, that is when we get closer to the Markovian DFS, it was shown \cite{orszag2} that the whole 
phenomenon of sudden death and revival disappears for the initial state $|\Psi_1\rangle$. Contrary to that, we see 
[for the initial state $|\Psi_1\rangle$ with $\epsilon=0.9$ 
in Fig.\ref{Mani1}(c)] clear sudden death and revival  for 
this range of $\epsilon$ as well in the {\it non-Markovian} case. 

\begin{figure}
\includegraphics[width=\linewidth]{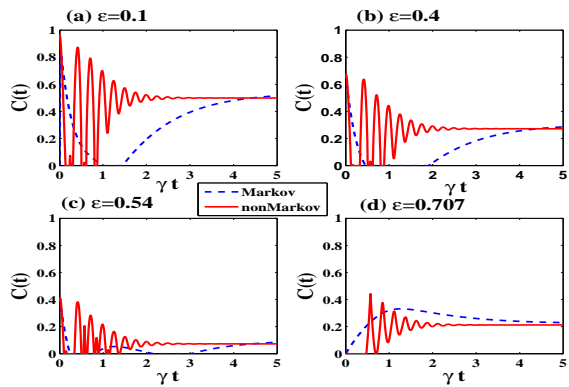}
\caption{(Color online) Non-Markovian (solid lines) and Markovian (dashed lines) 
time evolution of the concurrence for $|\Psi_2\rangle$ as an initial state with 
varying $\epsilon$: (a) $\epsilon=0.1$, (b) $\epsilon=0.4$, (c) $\epsilon=0.54$, and 
(d) $\epsilon=0.707$. The values of the other parameters are $r=0.31$, $\theta=0$, 
$\gamma=1$, and $\omega_c=\omega_0=1$. The number of ESD and revivals in the 
non-Markovian case increases. The state $|\phi_2\rangle$ ($|\Psi_2\rangle$ with 
$\epsilon=1$) remains decoherence free for both the Markovian and non-Markovian 
evolution.}
\label{Mani2}
\end{figure}

Fig.~\ref{Mani2} shows the dynamical behavior of the 
entanglement in terms of the concurrence for the initial state $|\Psi_2\rangle$. The state $|\Psi_2\rangle$ is a 
superposition of two states $|\phi_2\rangle$ (belonging to the Markovian DFS) and its orthogonal $|\phi_3\rangle$. 
We again vary $\epsilon$ between $0$ and $1$ for fixed values of the parameters $r=0.31$ and $\theta=0$. For the 
initial state $|\Psi_2\rangle$, we also see multiple cycles of death and revival of entanglement in the non-Markovian 
regime (in the parameter interval $0 \le \epsilon < 1/\sqrt{2}$) showing completely different behavior from its Markovian 
counterpart \cite{orszag2}. When $1/\sqrt{2} \le \epsilon < 1$, no sudden death was observed for the state $|\Psi_2\rangle$ 
in the Markov regime whereas we see clear sudden death in this case also for the non-Markovian case [see Fig.~\ref{Mani2}(d)].

\begin {figure}
\includegraphics[width=\linewidth]{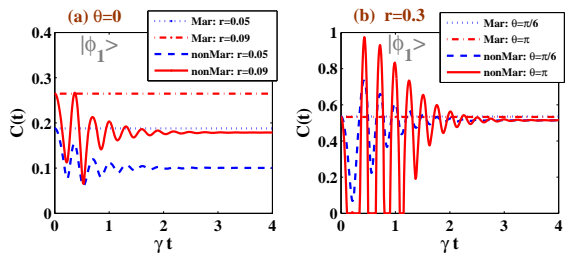}
\caption{(Color online) Non-Markovian entanglement dynamics for various Markovian
DFS states $|\phi_1\rangle$ with varying squeeze
parameters $r$ and $\theta$: (a)
$r=0.05$ and $r=0.09$ with $\theta=0$, (b) $\theta=\pi/6$ and $\theta=\pi$ with $r=0.3$.
The values of the other parameters are $\gamma=1$, and $\omega_c=\omega_0=1$.}
\label{prothom}
\end{figure}

Finally we emphasize again that the state $|\phi_1\rangle$ (the state $|\Psi_1\rangle$ with $\epsilon=1$) is not a 
decoherence free state for non-Markovian master equation (\ref{nonmarkov}) although it belongs to the DFS for Markovian 
master equation (\ref{markov}) [see Fig.~\ref{Mani1}(d)]. 
The Markovian and non-Markovian 
entanglement dynamics for the Markovian DFS states $|\phi_1\rangle$,
Eq.~(\ref{phi1}), with varying squeeze 
parameters $r$ and $\theta$ is shown in Fig.~\ref{prothom}.
We consider in Fig.~\ref{prothom} two separate cases (a) 
$r=0.05$, $r=0.09$ with $\theta=0$ and (b) $\theta=\pi/6$,
$\theta=\pi$ with $r=0.3$. 
We find that the initial entanglement  
and final asymptotic entanglement at large times (Markov and non-Markov) 
for the initial Markovian DFS states, Eq.~(\ref{phi1}), depend on $r$
not on $\theta$. 
One can observe from Fig.~\ref{prothom}(b) that when the squeeze
parameter $r$ is fixed and the 
phase $\theta$ is varied, the Markovian concurrence curves do not vary
and they remain the same and overlap. 
This is because when $r$ is varied, the values of $N$ and $M$ 
in Eq.~(\ref{phi1}) change considerably and so do the resultant Markovian
DFS states, while different values of $\theta$ change only the
relative phase between $|11\rangle$ and  $|00\rangle$ in the resultant
Markovian DFS states of Eq.~(\ref{phi1}). Nevertheless, one can see that these 
Markovian DFS states are not decoherence-free in the non-Markovian regime
and the characteristics of this non-Markovian oscillations of
concurrence depend  
on the squeeze parameters $r$ and $\theta$. On the other hand, the state 
$|\phi_2\rangle$ (the state $|\Psi_2\rangle$ with $\epsilon=1$) is decoherence 
free both for Markovian and non-Markovian master equations. This was
verified by obtaining a straight line $C(t)=1$, showing that the entanglement
remains constant for this state at zero temperature of the reservoir. 

\begin{figure}
\includegraphics[width=\linewidth]{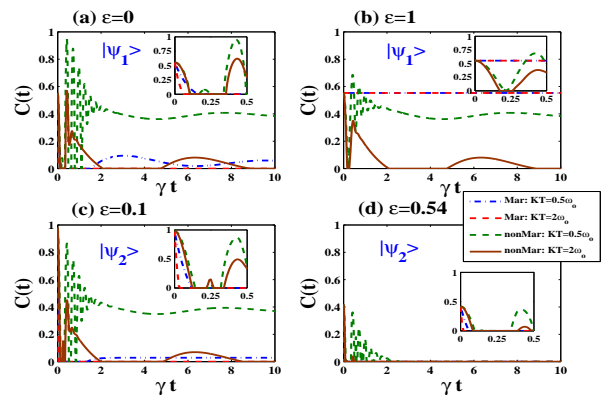}
\caption{(Color online) Non-Markovian and Markovian time evolution of
the concurrence at
finite temperatures for $|\Psi_1\rangle$ as an initial state with
(a) $\epsilon=0$ and  (b) $\epsilon=1$, and
for $|\Psi_2\rangle$ as an initial state with 
(c) $\epsilon=0.1$ and (d) $\epsilon=0.54$.
The values of the other parametrs are $r=0.31$, $\theta=0$, $\gamma=1$, and
$\omega_c=\omega_0=1$. The insets show the entanglement dynamics in the short-time region.
The non-Markovian entanglement oscillations gradually disappear as one
increases the temperature. 
The state $|\phi_2\rangle$ remains decoherence free both in the Markov
and non-Markov regime at any temperature of the reservoir.
But the state $|\phi_1\rangle$ ($|\Psi_1\rangle$ with 
$\epsilon=1$) that is decoherence free at any reservoir temperature 
in the Markovian case 
no longer remains decoherence free for the non-Markovian 
evolution.}
\label{Mani3}
\end{figure}

Next, we go to the finite-temperature case for which $N(\omega)$ and $M(\omega)$ are given by
Eqs.(\ref{Nw}) and (\ref{Mw}). To show the effect of temperature on the Markovian evolution 
let us focus on the entanglement dynamics of 
two specific initial states ($|\Psi_1\rangle$ with $\epsilon=0$, and $|\Psi_2\rangle$ with $\epsilon=0.1$).
From the zero-temperature Markovian dynamics of these two states [see Fig.~\ref{Mani1}(a) and Fig.~\ref{Mani2}(a)] 
we see that at long time the entanglement finally saturates to a finite value after which no death of entanglement 
occurs, whereas at finite temperatures (say, $KT > 2 \omega_0$) Markovian dynamics of these two states show a 
complete death of entanglement at long-time limit 
[see Figs.~\ref{Mani3}(a), \ref{Mani3}(c) and their insets].
From Fig.~\ref{Mani2}(a) we see for the zero-temperature case, 
Markovian entanglement sudden death occurs at relatively slower rate whereas in the finite-temperature case 
[Fig.~\ref{Mani3}(c)] ESD rate is much faster. The difference between 
the zero-temperature entanglement dynamics and 
the finite-temperature entanglement dynamics is more transparent in the non-Markovian case. The non-Markovian entanglement 
dynamics at zero temperature shows several death and revival cycles in short time limit while there
is only one or two death-revival cycles following a decay of entanglement for the finite temperature
(say, $KT > 2 \omega_0$) non-Markovian case. In the non-Markovian evolution (Fig.~\ref{Mani3}), as one gradually 
increases the temperature, the number of death and revival cycles of the concurrence reduces and the effect of temperature 
on concurrence in general is seen as decaying in nature (concurrence asymptotically reaches to its zero value at large times)  
with an exception to the DFS. That is in the long time limit, the non-Markovian entanglement saturates to a certain nonzero 
value at low temperature, while at high temperatures ($KT > 2 \omega_0$) the concurrence finally decays to zero in the long-time 
limit. Note also that the concurrence disappear more quickly at finite temperatures for both Markovian and non-Markovian 
cases. It is seen that the death time for the Markovian dynamics is always less compared to the non-Markovian dynamics at 
a given finite temperature. We have seen that the state $|\phi_2\rangle$ (the state $|\Psi_2\rangle$ with $\epsilon=1$) 
remains decoherence free both in the Markov and non-Markov regime at any finite temperature of the reservoir. This was 
verified by obtaining a straight line $C(t)=1$, showing that the entanglement remains constant for this state $|\phi_2\rangle$ 
at any arbitrary temperature of the reservoir. It is quite striking that the state $|\phi_1\rangle$ shows entanglement sudden 
death at finite temperature in the non-Markovian dynamics, althouth it remains a DFS at finite temperature for the Markovian
case [see Fig.~\ref{Mani3}(b)]. We have also checked numerically that non-Markovian oscillations of entanglement gradually 
diminishes as one increases $N$ defined in (\ref{Nw}), for fixed values of other parameters.

\section{Conclusion}\label{sec:conclusion}

In summary, we study the non-Markovian entanglement dynamics of two qubits in a common squeezed 
bath. We consider the initial two-qubit states which are very close to (as well as far from) the 
Markovian DFS for this system. We see (Fig.\ref{Mani1} and Fig.\ref{Mani2}) multiple cycles of 
entanglement sudden death and revival in the non-Markovian case, showing striking difference 
between the Markovian and non-Markovian entanglement dynamics. We also observe that a 
non-Markovian decoherence free state (for example, $|\phi_2\rangle$) remains decoherence 
free in the Markovian regime, but all the Markovian decoherence free states (for example, 
$|\phi_1\rangle$) are not necessarily decoherence free in the non-Markovian domain 
[Fig.\ref{Mani1}(d)]. Finally, we extend our result for the finite-temperature case 
where we see the non-Markovian entanglement oscillations gradually decreases as one 
increases the temperature. We found that the Markovian decoherence-free states remain 
invariant under finite temperatures [Fig.\ref{Mani3}(b)]. We also see from Fig.\ref{Mani3} 
that the finite temperature of the bath accelerates the phenomenon of entanglement sudden 
death for the non-decoherence-free entangled initial states. Interestingly, the state 
$|\phi_2\rangle$ is found to be decoherence free both in the Markov and non-Markov regime, 
and is also robust against finite bath temperatures. 
There were considerable number of 
papers in recent literature \cite{orszag1,orszag2,squeeze} dealing
with the squeezed reservoir. Also, there were 
proposals \cite{gardiner1,scully1,expt1,parkins1993,lutken,werlang,parkins2003} for physical 
realizations of environments that mimic or generate Markovian 
squeezed bath. One can suggest in future experimental schemes to test and observe various 
important issues related to non-Markovian dynamics by physically engineering the environment.
Hence, it will be interesting to see if our prediction of the non-Markovian effects can be 
verified in actual experiments.

\begin{acknowledgments}
We would like to acknowledge support from the National Science
Council, Taiwan, under Grant No. 97-2112-M-002-012-MY3, 
support from the Frontier and Innovative Research Program 
of the National Taiwan University under Grants No. 97R0066-65 and 
No. 97R0066-67,
and support from the focus group
program of the National Center for Theoretical Sciences, Taiwan.
We are grateful to the National Center for High-performance Computing, Taiwan, 
for computer time and facilities.
\end{acknowledgments}

\end{document}